\documentclass[12pt, epsfig]{article}
\usepackage{epsfig}
\usepackage{amssymb}
\usepackage{amsfonts}

\begin{document}


\newcommand{\beq}{\begin{equation}}
\newcommand{\eeq}{\end{equation}}
\newcommand{\bea}{\begin{eqnarray}}
\newcommand{\eea}{\end{eqnarray}}
\newcommand{\beqn}{\begin{eqnarray}}
\newcommand{\eeqn}{\end{eqnarray}}
\newcommand{\beas}{\begin{eqnarray*}}
\newcommand{\eeas}{\end{eqnarray*}}
\newcommand{\defi}{\stackrel{\rm def}{=}}
\newcommand{\non}{\nonumber}
\newcommand{\bquo}{\begin{quote}}
\newcommand{\enqu}{\end{quote}}
\newcommand{\p}{\partial}


\def\de{\partial}
\def\Tr{ \hbox{\rm Tr}}
\def\const{\hbox {\rm const.}}
\def\o{\over}
\def\im{\hbox{\rm Im}}
\def\re{\hbox{\rm Re}}
\def\bra{\langle}\def\ket{\rangle}
\def\Arg{\hbox {\rm Arg}}
\def\Re{\hbox {\rm Re}}
\def\Im{\hbox {\rm Im}}
\def\diag{\hbox{\rm diag}}

\def\stroke{\vrule height8pt width0.4pt depth-0.1pt}
\def\topfleck{\vrule height8pt width0.5pt depth-5.9pt}
\def\botfleck{\vrule height2pt width0.5pt depth0.1pt}
\def\Zmath{\vcenter{\hbox{\numbers\rlap{\rlap{Z}\kern 0.8pt\topfleck}\kern
2.2pt\rlap Z\kern 6pt\botfleck\kern 1pt}}}
\def\Qmath{\vcenter{\hbox{\upright\rlap{\rlap{Q}\kern
3.8pt\stroke}\phantom{Q}}}}
\def\Nmath{\vcenter{\hbox{\upright\rlap{I}\kern 1.7pt N}}}
\def\Cmath{\vcenter{\hbox{\upright\rlap{\rlap{C}\kern
3.8pt\stroke}\phantom{C}}}}
\def\Rmath{\vcenter{\hbox{\upright\rlap{I}\kern 1.7pt R}}}
\def\Z{\ifmmode\Zmath\else$\Zmath$\fi}
\def\Q{\ifmmode\Qmath\else$\Qmath$\fi}
\def\N{\ifmmode\Nmath\else$\Nmath$\fi}
\def\C{\ifmmode\Cmath\else$\Cmath$\fi}
\def\R{\ifmmode\Rmath\else$\Rmath$\fi}


\def\QATOPD#1#2#3#4{{#3 \atopwithdelims#1#2 #4}}
\def\stackunder#1#2{\mathrel{\mathop{#2}\limits_{#1}}}
\def\stackreb#1#2{\mathrel{\mathop{#2}\limits_{#1}}}
\def\Tr{{\rm Tr}}
\def\res{{\rm res}}
\def\Bf#1{\mbox{\boldmath $#1$}}
\def\balpha{{\Bf\alpha}}
\def\bbeta{{\Bf\beta}}
\def\bgamma{{\Bf\gamma}}
\def\bnu{{\Bf\nu}}
\def\bmu{{\Bf\mu}}
\def\bphi{{\Bf\phi}}
\def\bPhi{{\Bf\Phi}}
\def\bomega{{\Bf\omega}}
\def\blambda{{\Bf\lambda}}
\def\brho{{\Bf\rho}}
\def\bsigma{{\bfit\sigma}}
\def\bxi{{\Bf\xi}}
\def\bbeta{{\Bf\eta}}
\def\d{\partial}
\def\der#1#2{\frac{\d{#1}}{\d{#2}}}
\def\Im{{\rm Im}}
\def\Re{{\rm Re}}
\def\rank{{\rm rank}}
\def\diag{{\rm diag}}
\def\2{{1\over 2}}
\def\ntwo{${\cal N}=2\;$}
\def\4N{${\cal N}=4$}
\def\none{${\cal N}=1\;$}
\def\x{\stackrel{\otimes}{,}}
\def\beq{\begin{equation}}
\def\eeq{\end{equation}}
\def\ba{\beq\new\begin{array}{c}}
\def\ea{\end{array}\eeq}
\def\be{\ba}
\def\ee{\ea}
\def\stackreb#1#2{\mathrel{\mathop{#2}\limits_{#1}}}

\def\baselinestretch{1.0}

\begin{titlepage}

\begin{flushright}
FTPI-MINN-06-37\\
UMN-TH-2529/06\\
April 24, 2007
\end{flushright}

\vspace{0.5cm}

\begin{center}

{\Large \bf Low-Energy Limit of Yang--Mills\\[0.4mm]
with Massless Adjoint Quarks:\\[2.7mm]
Chiral Lagrangian and Skyrmions
}
\end{center}

\vspace{0.5cm}

\begin{center}
{\bf R. Auzzi  and \bf M.~Shifman}
\end {center}
\begin{center}

{\it  William I. Fine Theoretical Physics Institute,
University of Minnesota,
Minneapolis, MN 55455, USA}\\
\end{center}

\vspace{3mm}

\begin{abstract}

If the fundamental quarks of QCD are replaced by massless adjoint
quarks the pattern of the chiral symmetry breaking drastically changes compared
to the standard one. It becomes SU$(N_f)\to\,$SO$(N_f)$. 
While for $N_f=2$ the chiral Lagrangian describing
the ``pion" dynamics is well-known, this is not the case at $N_f>2$.
We outline a general strategy for deriving chiral Lagrangians for the coset spaces
$\mathcal{M}_k={\rm SU}(k)/{\rm SO}(k)$,
and study in detail the case of $N_f=k=3$. We obtain two-
and four-derivatives terms in the chiral Lagrangian 
on the coset space $\mathcal{M}_3$ = SU(3)/SO(3), as well as 
the Wess--Zumino--Novikov--Witten term,
in terms of an explicit parameterization of the quotient manifold.
Then we discuss stable topological solitons supported
by this Lagrangian. Aspects
of relevant topological considerations scattered in the
literature are reviewed.
The same analysis applies to  SO$(N)$ gauge theories with $N_f$
Weyl flavors in the vector representation. 

\end{abstract}

\end{titlepage}

\section{Introduction}
\label{intro}

Recently a nontrivial large-$N$ equivalence
between bosonic subsectors of  different gauge theories
has been established \cite{asv} (for a review see \cite{asv-rev}). This planar
equivalence connects, in particular,
the Yang--Mills theory with $N_f$ Dirac fermions in the two-index
symmetric (or antisymmetric)   representation of color SU$(N)$ on the one side,
with the  theory with $N_f$ Weyl quarks in the adjoint representation on the other side.

If the number of flavors $N_f >1$, both theories under consideration have a
chiral symmetry which is spontaneously broken. The pattern of the chiral
symmetry breaking ($\chi$SB) is different \cite{mac,triangle,rm}. For $N_f$
Dirac fermions in the two-index (anti)symmetric representation the pattern of 
$\chi$SB is identical to that of QCD, namely,
\begin{equation}
\mathrm{SU}(N_f)_L\times \mathrm{SU}(N_f)_R \to \mathrm{SU}(N_f)_V
\label{vsez}
\end{equation}
On the other hand, in the
SU$(N)$ gauge theories with $N_f$ {\em Weyl fermions in
the adjoint} representation we have the following $\chi$SB
 pattern,\footnote{To ensure the very existence of the global chiral symmetry on the
one hand, and to keep the microscopic theory asymptotically free on the
other, we must assume
that  $ 2\leq N_f\leq 5\,. $
A more exact version of Eq.~(\ref{patmi})
is
$\mathrm{SU}(k)\times \mathbb{Z}_{2N}\to \mathrm{SO}(k)\times 
\mathbb{Z}_{2} $
where the discrete factors are the remnants of the anomalous singlet axial
U(1); they play no role in what follows.}
\beq
{\rm SU} (N_f)\rightarrow {\rm SO}(N_f)\,.
\label{doptue}
\eeq
Thus, in this case the low-energy effective  theory 
 is a sigma model on the target space
 \beq
 \mathcal{M}_k= {\rm SU}(k)/{\rm SO}(k) 
 \label{patmi}
\eeq
with $k=N_f$. This effective theory describes the interactions of 
the Goldstone bosons of the theory, the  ``pions."  
[Let us note in passing that the
same sigma model emerges in  SO$(N)$ gauge theories with $N_f$ Weyl 
fermions in the {\em vector} representation. In this case 
for large enough $N$  there is no
upper bound on $N_f$.] 

For two adjoint flavors, $ \mathcal{M}_2= {\rm SU}(2)/{\rm SO}(2) = S^2$. 
The corresponding sigma model is a well-studied O(3) sigma model \cite{Polyakov}. With a four-derivatives term included it goes under the
name of the Skyrme--Faddeev model (or, sometimes, the Faddeev--Hopf model)
\cite{Faddeev}. Solitons in this model are intriguingly interesting because of their
knotted structure.  They are known as Hopf solitons and were extensively studied
\cite{Faddeev,bs} within the framework of a ``glueball hypothesis" \cite{Faddeev} according to which the Hopf solitons may be relevant to the description of glueball 
states in pure Yang--Mills theory. The fact that they are certainly relevant
in the studies of solitons built from pions was noted in \cite{BoShi} where a detailed analysis of the $N_f=2$ case is presented. In application to chiral Lagrangians, it is natural to refer to these solitons as Hopf Skyrmions. At large $N$ the quasi classical consideration of the Skyrmions\,\footnote{The Skyrmions were introduced in particle physics long before QCD \cite{skyrme}.}  becomes theoretically justified \cite{witten,witten2}; therefore, these solitons should be in  one-to-one correspondence
with certain hadronic states from the spectrum of the given microscopic theory
(see Ref. \cite{bo} for a discussion of this problem in Yang--Mills with two-index
(anti)symmetric matter). 

As was mentioned, the two-flavor case is singled out by the fact that
in this problem the effective low-energy Lagrangian is known,
so that its analysis, as well as that of solitons it supports, can be carried out in 
more or less explicit manner, through a combination of analytic and numeric 
methods (see \cite{BoShi}). At the same time,
to the best of our knowledge,
sigma models on the target spaces (\ref{patmi})
with $k=3,4$ and $5$ have not been studied in the literature so far.
In this work we fill the gap.
First, we outline general considerations referring to three,
four and five flavors. Then we derive, in an explicit form, the chiral Lagrangian
for the sigma model on $\mathcal{M}_3$. 
We discuss its features in much detail. In particular, we discuss solitons
in this model, and how they match the Hopf Skyrmions of the
$N_f=2$ model if one ascribes a large mass term to the third flavor.

The topology of the target space gives us 
information about the solitons in the model. 
The second homotopy group, $\pi_2(\mathcal{M}_k)$, is relevant for
the spectrum of the flux tubes.\footnote{As was pointed out in 
Ref.~\cite{benson-saadi} the Skyrme
  model with just the quadratic and  quartic terms
  exhibits  size instability for all vortices:
  under a spatial rescaling $r\rightarrow \lambda r$
  the quadratic contribution stays invariant and
  the quartic one rescales by a factor $\lambda^2$.
Therefore,  the energy is minimized at infinite size.
  As discussed in Ref. \cite{benson-saadi},
  this divergence can be eliminated, for example, by giving 
a bare mass  to the quarks, 
which explicitly breaks the flavor symmetry and
  induces a potential term on the target space of the sigma model under consideration.}
On the other hand, $\pi_3(\mathcal{M}_k)$ gives us the spectrum of the particle-like solitons (Skyrmions).  Moreover, $\pi_4(\mathcal{M}_k)$ and $\pi_5(\mathcal{M}_k)$ are relevant
for the introduction of the Wess--Zumino--Novikov--Witten (WZNW) 
\cite{wznw} term
which, in the case of QCD,
tells us how to quantize  the Skyrmion, i.e. whether it becomes
a fermion or a boson upon quantization \cite{witten,witten2}.
The relevant homotopy groups are shown in Table~1.

\begin{table}[h]     
\begin{center}    
\begin{tabular}  {|l|l|l|l|l|l|} \hline $k$ & dim $\mathcal{M}_k$ & $\pi_2$ & $\pi_3$ & $\pi_4$ & $\pi_5$\\[2mm]
\hline
$2$ & $2$ & $\mathbb{Z}$ & $\mathbb{Z}$ & $\mathbb{Z}_2$ & $\mathbb{Z}_2$ 
\\[2mm]
$3$ & $5$ & $\mathbb{Z}_2$ & $\mathbb{Z}_4$ & $1$ & $\mathbb{Z}\otimes\mathbb{Z}_2$ \\[2mm]
$4$ & $9$ &  $\mathbb{Z}_2$ & $\mathbb{Z}_2$  & $\mathbb{Z}$ & $\mathbb{Z}\otimes \mathbb{Z}_2
\otimes \mathbb{Z}_2$ \\[2mm]
 $5$ & $14$ & $\mathbb{Z}_2$ & $\mathbb{Z}_2$  & $1$ & $\mathbb{Z}\otimes \mathbb{Z}_2$\\[2mm]
 $k>5$ & $\frac{k^2+k-2}{2}$ & $\mathbb{Z}_2$ & $\mathbb{Z}_2$  & $1$ & $\mathbb{Z}$\\[2mm]
\hline
\end{tabular}   
\caption{\footnotesize Some homotopy groups 
 for the manifolds $\mathcal{M}_k$ (see Refs. 
 \cite{witten,rigas}). 
 The relevant exact sequences are discussed in Appendix~A.
 The WZNW term cannot be introduced for $k=2,4$
 because $\pi_4(\mathcal{M}_k)$ is nontrivial.
 On the other hand, the $\mathbb{Z}_2$ factors in $\pi_5(\mathcal{M}_{3,5})$
present no topological obstruction for the WZNW term.}     
\label{topologia}  
\end{center}
\end{table}

For $k=2$ the target space $\mathcal{M}_2$ reduces to a two-dimensional sphere $S^2$.
The corresponding sigma model supports flux tubes due
to the fact that  $\pi_2(\mathcal{M}_2)=\mathbb{Z}$,
which are classified by integer topological numbers.

Since $\pi_3(\mathcal{M}_2)=\mathbb{Z}$,
the Hopf Skyrmions are also classified by  integers 
 (these solitons can be understood as 
twisted flux tubes; mathematically this can 
be shown by using the Hopf fibration, which gives us
the first topologically nontrivial map between $S^3$ and $S^2$). 

Furthermore, 
we have  $\pi_4(S^2)=\mathbb{Z}_2$, implying that
 it is possible to quantize the Hopf Skyrmions
 both as  bosons or as  fermions \cite{fr,krusch,su1}. There is no WZNW term for two flavors. In order to have the  WZNW term, the target manifold of the sigma model
in question must have dimension  five or larger.

We will dwell on  the $N_f=3$ case. We will introduce
an explicit parameterization of the coset space $\mathcal{M}_3={\rm SU}(3)/
{\rm SO}(3)$
 and construct an explicit Lagrangian  for this sigma model,
 including the quadratic and quartic terms. 
It is shown that the homotopy class relevant for the $\mathbb{Z}_2$
vortices and for the $\mathbb{Z}_4$ Skyrmions supported by this Lagrangian
can be obtained by an embedding of the  corresponding homotopy class
from $\mathcal{M}_2$. The WZNW term will be  calculated. 
We will show that it is proportional to the 5-volume form on $\mathcal{M}_3$.

The organization of the paper is as follows.
In Sect.~\ref{sectwo}
we outline a general formalism allowing one to construct
sigma models on $G/H$ manifolds. We review the application
of this formalism to the SU(2)/SO(2) case;
as a warm-up exercise we
derive in this formalism the chiral
Lagrangian of the O(3) Skyrme-Faddeev model.
In Sect.~\ref{secthree} 
we apply it to the  $k=3$ case. We introduce
explicit coordinates and then obtain the metric on
$\mathcal{M}_3={\rm SU}(3)/{\rm SO}(3)$.
The two-derivative part of the Lagrangian is presented in Sect.~\ref{secthree},
while the four-derivative part in Appendix B.
Topological aspects relevant to various solitonic configurations 
in the ${\rm SU}(3)/{\rm SO}(3)$ sigma model are discussed
in Sect.~\ref{secfour}. The WZNW term on
$\mathcal{M}_3$ is calculated in Sect.~\ref{secfive}.
Appendix A presents the exact sequences for some homotopy groups
used in the paper.

\section{General considerations}
\label{sectwo}

To refresh memory, it is convenient to
start from the well-known case of QCD with $N_f$ Dirac quarks. Then the
 Lagrangian of the
 Skyrme model includes the following two- and four-derivatives terms:
\beqn
\mathcal{L} &=&  \frac{F_\pi^2}{4} \mathcal{L}_2+ \frac{1}{e^2} \mathcal{L}_4 \nonumber\\[2mm]
&=& \frac{F_\pi^2}{4} \Tr \, (\partial_\mu U \partial^\mu U^\dagger)+
\frac{1}{e^2} \Tr \, [ (\partial_\mu U) U^\dagger,(\partial_\nu U) U^\dagger ]^2, 
\eeqn
where the matrix $U$ is an element of the SU$(N_f)$ group,
and $F_\pi$ and $e$ are constants.
The subscript $\pi$ will be omitted hereafter. 
The two-derivatives term is just the kinetic 
term of the Goldstone bosons of the theory;
mathematically it is the metric of the target manifold.
The four-derivatives term is needed in order to stabilize
the particle-like solutions, which otherwise would tend
to shrink to zero size. The coset space corresponding to (\ref{vsez}) is a groups space
itself.

In the generic case of the group quotient $G/H$, 
a general prescription for obtaining two-derivatives terms
was given long ago in Ref. \cite{ccwz}.
This issue has been recently discussed anew in Ref. \cite{koshkin}
in a slightly modified perspective pertinent to the Faddeev--Skyrme models.
Following the formalism of the latter paper, we get for
the two-derivatives term
\beq 
\mathcal{L}_2=\Tr \, \left(
P_{h\bot}(U^\dagger \partial_\mu U)\cdot P_{h\bot}(U^\dagger \partial_\mu U) 
\right),
\label{la2} 
\eeq
where $P_{h\bot}$ is the projection in the Lie algebra of $G$ on the space orthogonal
to the Lie algebra of $H$, which we call $h$. The construction of $P_{h\bot}$
will be discussed momentarily.
Analogously, the four derivatives term is
\beq 
\mathcal{L}_4=\Tr \left[P_{h\bot}(U^\dagger \partial_\mu U), P_{h\bot}(U^\dagger \partial_\nu U) \right]^2 \,. 
\label{la4} 
\eeq

As a warm-up exercise let us discuss first the Faddeev--Skyrme model,
in which $G$=SU(2) and $H$=SO(2),
and the explicit form of the Lagrangian is well known.
The quotient can be parameterized using the
matrix exponential of the  SU(2)  generators which
are not in the chosen $H={\rm SO}(2)=U(1)$.
Let us assume that  the $U(1)$
factor is generated by the second the Pauli matrix,
\beq
e^{i\, \sigma_2 \,t}\,,\qquad
\sigma_2=\pmatrix{ 0  &  -i 
  \cr i   &  0         } \,.
 \eeq
Then  these generators of $G/H$ are the symmetric
self-adjoint two-by-two matrices.  Any such element can be parametrized as

\beq 
U=\exp (i \, V\cdot A\cdot V^\dagger), 
\label{eight}
\eeq
where $A$ is the diagonal matrix,
\beq 
A=\pmatrix{  + \theta/2     &  0  
  \cr  0   &  - \theta/2         },
  \label{nine}
  \eeq
  and 
\beq V=\pmatrix{  \cos \alpha/2     \rule{0mm}{6mm}  &  -\sin \alpha /2  
  \cr  \sin \alpha /2   \rule{0mm}{6mm}   &  \cos \alpha/2         }.
  \label{ten}
  \eeq
With this parametrization we recover the standard $S^2$,
provided
\beq
0\leq \theta \leq \pi\,,\quad0 \leq \alpha \leq 2 \pi\,.
\label{eleven}
\eeq
Indeed, the
projection $P_{h\bot}(T)$  defined on the
Lie algebra of  SU(2)  is given by
\beq 
P_{h\bot}(T)=
T- \frac{1}{2}\,  \sigma_2\, \Tr\, (T\cdot \sigma_2)\,.
\label{elevenp}
\eeq
Then we obtain Eq.~(\ref{eight}) and, 
from Eq.~(\ref{la2}) we arrive at
the two-derivatives term presenting  the standard on $S^2$, 
\beqn
&& \frac{1}{2} \left[(\partial_\mu \theta)^2 +\sin^2 \theta \, (\partial_\mu \alpha)^2\right]= \frac{1}{2}(\partial_\mu \vec{n})^2\,,\nonumber\\[2mm]
 &&  \vec{n}\cdot\vec{n} =1\,.
 \eeqn
Furthermore,  the four-derivatives term is recovered from Eq.~(\ref{la4}),
\beq 
\frac{1}{2} \sin^2 \theta\,
(\partial_\mu \theta \, \partial_\nu \alpha-
\partial_\nu \theta \, \partial_\mu \alpha)^2\,,
\label{fourteen}
\eeq
which identically reduces to
\beq
\frac{1}{2}\left(\partial_\mu \vec{n} \wedge \partial_\nu \vec{n}\right)^2\,.
\label{fifteen}
\eeq

\section{An explicit Lagrangian for \boldmath{$k=3$}}
\label{secthree}

We can proceed in a way analogous to 
what we have just done for the $k=2$ case.
We parameterize the quotient using the matrix exponential of the
generators of SU(3) which are not in SO(3).
These generators are the symmetric $3 \times 3$ matrices. It is always possible to diagonalize a symmetric matrix in  an orthogonal basis. We introduce the parameters 
$\theta,\eta$ for the eigenvalues of the matrix and the parameters 
$\alpha,\beta,\gamma$ as the Euler angles for
the transformation which brings the generic symmetric matrix in diagonal form.
The angular range of each of the five parameters mentioned above will be determined
using the SO(3) quotient relations. 

The parameterization we use is as follows:
\beq 
U=\exp (\, i  \,  V\cdot A \cdot V^\dagger) 
\label{sixteen}
\eeq
where 
\beq 
A=\frac{1}{2} \pmatrix{  \eta/\sqrt{3} + \theta     \rule{0mm}{6mm}&  0   &    0
  \cr  0   & \eta/\sqrt{3} - \theta    \rule{0mm}{6mm}   & 0
\cr  0    & 0  \rule{0mm}{6mm}   & -2\eta/\sqrt{3}     }
\label{seventeen}
\eeq
and $V$ is an SO(3) matrix parameterized by three Euler
angles $\alpha,\beta,\gamma$,
{\footnotesize 
$$
V=\pmatrix{ \cos \frac{\alpha}{2} \cos \frac{\gamma}{2}-\cos \frac{\beta}{2} \sin \frac{\alpha}{2}
 \sin \frac{\gamma}{2}\rule{0mm}{8mm}
   & - \sin \frac{\alpha}{2} \cos \frac{\gamma}{2}-\cos \frac{\alpha}{2} \cos \frac{\beta}{2} \sin \frac{\gamma}{2}   \rule{0mm}{8mm}
   &    \sin \frac{\beta}{2} \sin \frac{\gamma}{2}
  \cr \cos \frac{\alpha}{2} \sin \frac{\gamma}{2}+\cos \frac{\beta}{2}
   \sin \frac{\alpha}{2} \cos \frac{\gamma}{2}\rule{0mm}{8mm}
    & -\sin \frac{\alpha}{2} \sin \frac{\gamma}{2} + \cos \frac{\alpha}{2} \cos \frac{\beta}{2} 
    \cos \frac{\gamma}{2} \rule{0mm}{8mm}     
    & -\cos \frac{\gamma}{2} \sin \frac{\beta}{2}
\cr  \sin \frac{\alpha}{2} \sin \frac{\beta}{2}  
\rule{0mm}{8mm} & \cos \frac{\alpha}{2} \sin \frac{\beta}{2} 
  & \cos \frac{\beta}{2}    
  }
$$
\beq\mbox{}
  \label{eighteen}
\eeq}
The angle variation range for the $\theta$ is 
\beq  
0\leq \theta \leq \pi\,.
\label{ninteen}
 \eeq 
The range for $\theta$ comes from
 the following equivalence which holds modulo SO(3) conjugation:
\beq 
\theta \rightarrow 2 \pi - \theta, \quad \alpha \rightarrow \alpha \pm \pi\,.
\label{twenty}
\eeq
In other words,
\beq 
U_{\theta,\alpha}\cdot (U_{2 \pi-\theta,\alpha \pm \pi})^{-1}\in {\rm SO}(3)
\,.
\label{twenty1}
\eeq
In addition, the action of $\alpha$ rotations modulo SO(3) is trivial at $\theta=\pi$.
The range for $\eta$ is
\beq 
-\frac{\theta}{\sqrt{3}} \leq \eta \leq \frac{\theta}{\sqrt{3}}  \rule{0mm}{8mm}  \,.
\label{twenty2}
  \eeq
This is due to the fact that we do not have to double-count different eigenvalue
orderings
(we can make an arbitrary permutation of the diagonal elements by applying 
 a combination of $\alpha=\pi$, $\beta=\pi$ and $\gamma=\pi$ rotations).
At $\eta=\pm \theta / \sqrt{3}$ we observe that two of the three elements are degenerate.

\begin{figure}[ht]
\begin{center}
\leavevmode
\epsfxsize 9  cm
\epsffile{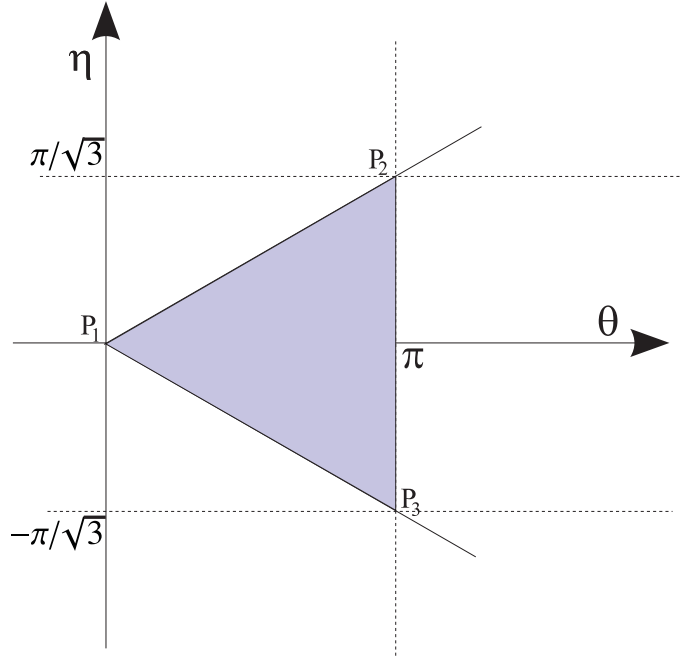}
\end{center}
\caption{\footnotesize Range of variation for $\theta$ and $\eta$.
At the triangle vertices the action of the Euler rotations
is trivial (they correspond to points on the $\mathcal{M}_3$).
These three points are in correspondence with the center $\mathbb{Z}_3$
of SU(3). Any point on the sides of the triangle correspond 
to a two-dimensional submanifold,
while any point in the interior correspond to a three-dimensional submanifold. 
}
\label{theta-eta}
\end{figure}

Finally, the range of variation for
the Euler parameters is
\beq
  0\leq \alpha \leq  2  \pi,
  \quad
  0\leq \beta \leq 2 \pi,
  \quad
  0\leq \gamma \leq 2 \pi\,.
  \label{extr}
  \eeq
These ranges come from the following three distinct invariances for the matrix $U$:
\beqn 
&&\alpha \rightarrow \alpha+ 2 \pi, 
\nonumber\\[2mm]
&& \alpha \rightarrow 2 \pi -\alpha,
\quad \beta \rightarrow \beta+2 \pi, \nonumber\\[2mm]
&& \alpha \rightarrow 2 \pi -\alpha, 
\quad
\beta \rightarrow 2 \pi -\beta , 
\quad
\gamma \rightarrow \gamma+2 \pi\,. 
\eeqn

The Lie algebra of $H=$SO(3) is generated by  three Gell-Mann matrices,
$\lambda_2$, $\lambda_5$ and $\lambda_7$,
\beq
\lambda_2=\pmatrix{ 0     &  -i   &    0
  \cr  i   & 0      & 0
\cr  0    & 0     & 0     }, \,\,\,
\lambda_5=\pmatrix{ 0     &  0  &    -i
  \cr  0  & 0      & 0
\cr  i    & 0     & 0     }, \,\,\, 
\lambda_7=\pmatrix{ 0     &  0  &    0
  \cr  0  & 0      & -i
\cr  0    & i     & 0     }.
\rule{0mm}{1.3cm}
\eeq
\mbox{}
\rule{0mm}{0.7cm}
\noindent
The projector $P_{h \bot}(T)$ is
\beq 
P_{h \bot}(T)=T- \frac{1}{2}
\left\{ \lambda_2 \Tr\, (T\cdot \lambda_2) +
\lambda_5 \Tr\, (T\cdot\lambda_5)+ \lambda_7 \Tr\, (T\cdot\lambda_7)\right\}.  
\label{twenty6}
\eeq

The two-derivatives term can be obtained upon substituting the parametrization
(\ref{sixteen}) -- (\ref{eighteen}) in the general equation  (\ref{la2}), 
\beqn  
&&
\mathcal{L}_2= \frac{1}{4} \left[ 2(\partial_\mu \theta)^2+2(\partial_\mu \eta)^2
+2 \sin^2  \theta (\partial_\mu \alpha)^2 \right.  \nonumber\\[2mm]
&& +(1-\cos \sqrt{3} \eta \cos \theta-
\cos \alpha \sin  \sqrt{3} \eta \sin  \theta) 
(\partial_\mu \beta)^2\nonumber\\[2mm]
&& 
 +\frac{1}{2}(\partial_\mu \gamma)^2 (2-
(1+\cos \beta) \cos^2 \theta-2 \cos \sqrt{3} \eta \cos \theta \sin^2 \frac{\beta}{2}
\nonumber\\[2mm]
&&
 +2 \cos \alpha \sin^2 \frac{\beta}{2} \sin \sqrt{3} \eta \sin \theta+\sin^2 \theta+
\cos \beta \sin^2 \theta
 )\nonumber\\[2mm]
 &&
 \left. +(4 \cos \frac{\beta}{2} \sin^2  \theta)
(\partial_\mu \alpha)(\partial_\mu \gamma)-
(2 \sin  \alpha \sin \frac{\beta}{2} \sin  \sqrt{3} \eta \sin  \theta)
(\partial_\mu \beta)(\partial_\mu \gamma)\right].
\nonumber\\[2mm]
\label{metric}
\eeqn

As a nontrivial check we can compute from this metric the 
 the scalar curvature. We find that it is constant
 as is required for the  symmetric space,
 \beq 
 r=15\, .
 \label{twenty8}
 \eeq
Moreover, the Ricci tensor is proportional to the metric 
($\mathcal{M}_3$ is an
Einstein manifold,  as for many other coset spaces),
\beq
R_{ab}=3 g_{ab}\,,
\eeq
where $g_{ab}$ is just the metric in Eq.~(\ref{metric})
written in the tensorial form. 

The four-derivatives term can be computed from Eq.~(\ref{la4}); the result is quite bulky.
The explicit expression for the four-derivatives term is given in Appendix B.


\section{Topology and Solitons}
\label{secfour}

\subsection{Topology of the sections at constant $(\theta,\eta)$}

Let us discuss Fig.~\ref{theta-eta} in some detail. For every fixed value of 
$(\theta,\eta)$
we have a submanifold  $\mathcal{R}(\theta, \eta)$. 
First of all let us consider  the topology of
 $\mathcal{R}(\theta, \eta)$ for a generic value inside the triangle in 
 Fig.~\ref{theta-eta},
 $$
 -\theta/\sqrt{3}< \eta < \theta/\sqrt{3}\quad{\rm  and}\quad0< \theta < \pi\,. 
 $$
Each of these submanifolds is parameterized by a generic  SO(3) 
rotation with the Euler angles $\alpha,\beta,\gamma$. 

There is a subtle
point, however.  Some of these SO(3) elements have a trivial action. 
These element constitute a $\mathbb{Z}_2 \otimes \mathbb{Z}_2$ subgroup of
SO(3), let us call it $\mathcal{A}$,
\beq
\mathcal{A} =\{ 1, \, a, \, b, \, a\cdot b\}\,,
\eeq
where
\beq 
a= \pmatrix{  -1    &  0   &    0
  \cr  0   & -1      & 0
\cr  0    & 0     & 1 }, \,\,\, b= \pmatrix{  1    &  0   &    0
  \cr  0   & -1      & 0
\cr  0    & 0     & -1 }, \,\,\,
a \cdot b=\pmatrix{  -1    &  0   &    0
  \cr  0   & 1      & 0
\cr  0    & 0     & -1 }\,.
\eeq
It is not difficult to check that
 \beq 
 a^2=b^2=(a\cdot b)^2=1\,.
 \eeq
 From expressions above it is rather obvious that $\mathcal{A}$
 is a subgroup of SO(3). 
 
It is well known that SU(2) and SO(3) differ by the center element $\mathbb{Z}_2$,
\beq 
{\rm SO}(3)={\rm SU}(2)/\mathbb{Z}_2\,.
\label{ce}
\eeq 
It is convenient to introduce the projection operator
$\mathcal{P}$,
\beq
{\rm SU}(2)\,  \stackrel{\cal P}{\longrightarrow} \, {\rm SO}(3)\,.
\label{34}
\eeq
Now, we will need to build an eight-element subgroup $\tilde\mathcal{A}$ of SU(2)
 which is in the same relation to
$\mathcal{A}$ as in Eq. (\ref{34}), namely,
\beq
\tilde\mathcal{A} = {\cal P}^{-1}\, \mathcal{A}\,.
\eeq
The eight elements of the subgroup $\tilde\mathcal{A}$ are as follows:
let us call the $\mathbb{Z}_2$ center element in Eq. (\ref{ce}) as $\tilde{c}$. 
Moreover,
\beq 
\tilde{a}=\exp(i \pi \sigma_3/2), \qquad \tilde{b}=\exp(i \pi \sigma_1/2)\, .
\label{36}
\eeq
Then
 \beqn
\tilde\mathcal{A} =&& \Big\{
  1, \,\,\, \tilde{a}, \,\,\, \tilde{a}^2=\tilde{c}, \,\,\, \tilde{a}^3, \,\,\, 
 \nonumber\\[2mm]
&&  \tilde{a}\tilde{b}, \,\,\, \tilde{a}^2 \tilde{b},
     \,\,\, \tilde{a}^3 \tilde{b}, \,\,\, \tilde{a}^4 \tilde{b}=\tilde{b}\,\Big\}\,.  
      \label{pigreco}
\eeqn
This is a subgroup of SU(2) 
isomorphic to the dihedral group $D_4$, 
$$
\tilde\mathcal{A}\sim D_4\,.
$$
The group $D_4$ has 3 possible $\mathbb{Z}_4$ subgroups, each of
them generated by powers of $\tilde{a}$, $\tilde{b}$, $\tilde{a} \tilde{b}$.   
The  conclusion we arrive at is
 \beq 
 \mathcal{R}(\theta, \eta)={\rm SU}(2)/D_4\, ,
 \label{38}
 \eeq  
which entails
  \beq 
  \pi_1(\mathcal{R}(\theta, \eta))=D_4\,.
  \label{39}
  \eeq 
Equation (\ref{39})
is due to the fact that  SU(2)  is simply connected.

\vspace{2mm}

Now let us consider ``degenerate"  values of $(\eta,\theta)$  on a side
 of the triangle ($\eta=\pm \theta/\sqrt{3}$ or $\theta=\pi$)
 in Fig.~\ref{theta-eta}. In such points we have that the SO(3)
 group is degenerates into SO$(2) \times \mathbb{Z}_2$. For example, 
 at $\eta=-\theta/\sqrt{3}$,
the SO(2) subgroup is generated by 
\beq 
\exp \left\{ i \pmatrix{  0    &  0   &    1
  \cr  0   & 0      & 0
\cr  -1   & 0     & 0 } t\right\},
\label{40}
\eeq 
and the $\mathbb{Z}_2$ element is
\beq 
b= \pmatrix{  1    &  0   &    0
  \cr  0   & -1      & 0
\cr  0    & 0     & -1 } . 
\label{41}
\eeq  
 We conclude that on the three segments $\eta=\theta/\sqrt{3}$,
 $\eta=-\theta/\sqrt{3}$ and $\theta=\pi$
 \beq 
 \mathcal{R}(\theta, \eta)=\frac{{\rm SO}(3)/{\rm SO}(2)}{\mathbb{Z}_2}=S^2/\mathbb{Z}_2=\mathbb{PR}^2,
 \eeq  
which implies, of course, 
  \beq 
  \pi_1(\mathcal{R}(\theta, \eta))=\mathbb{Z}_2\, .
  \label{43}
  \eeq 

If, in consideration of the $\mathcal{R}(\theta, \eta)$ section
we continuously move from a point  in the internal part of the triangle to
a point on one of its three sides, we have that a $\mathbb{Z}_4$ subgroup of
the fundamental group $D_4$ becomes trivial. 
We have that a different $\mathbb{Z}_4$ subgroup becomes trivial on each of the sides of the triangle, namely,
 \beqn
&& \eta=\theta/\sqrt{3} \rightarrow (1,\tilde{b},\tilde{b}^2,\tilde{b}^3)\,,
 \nonumber\\[2mm]
&& \eta=-\theta/\sqrt{3} \rightarrow (1,\tilde{a}\tilde{b},(\tilde{a}\tilde{b})^2,
(\tilde{a}\tilde{b})^3)\,, \nonumber\\[2mm]
&&  \theta=\pi \rightarrow (1,\tilde{a},\tilde{a}^2,\tilde{a}^3)\,. 
\label{44}
\eeqn   

Finally, if we consider  the vertices $P_{1,2,3}$ of the triangle,
the action of the Euler rotations modulo the unbroken SO(3) is trivial.
Therefore, in correspondence with these three values,  we find that 
 $ \mathcal{R}(P_{1,2,3})$ is a point.

\subsection{Homotopy group generators}

After this discussion we are ready to elucidate how to build explicitly
the 2- and 3-cycles in our parameterization. The vortex soliton
will wrap on a nontrivial 2-cycle while the Skyrmion will wrap on a
3-cycle, so this discussion is important for understanding of how to build
the solitons in the theory at hand. 

Let us start with the 2-cycle. From Ref. \cite{witten} we know 
that $$\pi_2(\mathcal{M}_3)=\mathbb{Z}_2\,.$$
Hence,  the problem is to identify the homotopy class of the only topologically nontrivial
map from $S^2$ onto $\mathcal{M}_3$. 

Let us denote $(\theta_{\rm s},\phi_{\rm s})$ the standard coordinates on $S^2$.
We then can build this nontrivial map in the following way:
we map the north pole of $S^2$ onto $P_1$, and the south pole onto $P_2$. 
We  map
the $\theta_{\rm s}$ coordinate of the sphere along the 
$\eta=-\theta/\sqrt{3}$ line, with
the relation $\theta_{\rm s}=\theta$. 
The $\phi_{\rm s}$ coordinate, on the other hand,
is mapped continuously onto a representative of the nontrivial 1-cycle of 
$$\pi_1(\mathcal{R}(\theta, \eta=-\theta/\sqrt{3}))=\mathbb{Z}_2\,.$$ 
For example, $\beta,\gamma=0$ and $\alpha=\phi_{\rm s}$. 

There is no  way
to shrink this map to a point.  It is possible, say, to   continuously transform the map 
from the segment $P_1P_2$ to  $P_1P_3$ or to $P_2P_3$, but it is impossible to shrink the map to trivial in this way.  
Also, if we compose this map twice, as in the definition of $\pi_2$, we find
a topologically trivial map. 

This map is also homotopic to the map
\[ \theta=\theta_{\rm s}, \,\,\, \alpha=\phi_{\rm s}, \]
with $\eta=\beta=\gamma=0$.
The image of this map is
in the $\mathcal{M}_2$ submanifold of $\mathcal{M}_3$
defined by the constraint $\eta=\beta=\gamma=0$;
the homotopy class corresponds to the vortex with the minimal winding in 
$\mathcal{M}_2$. 
This shows that if we embed the minimal winding vortex  
of the Faddeev--Skyrme model in  $\mathcal{M}_3$,  we obtain a representative
of the homotopy class of the $\mathbb{Z}_2$ vortex. On the other hand,
the vortices with nonminimal winding are unstable if embedded in $\mathcal{M}_3$:
 the ones with the even winding number will decay to the topologically trivial configuration and the
ones with the odd winding number to the $\mathbb{Z}_2$ minimal vortex.

From the exact sequence of the homotopy group of a fiber bundle (discussed
in Appendix A), we know 
not only that $$\pi_3(\mathcal{M}_3)=\mathbb{Z}_4\,,$$ but, in addition,  that
the   elements of  $\mathbb{Z}_4$ are the projection modulo 4 of 
$$\pi_3({\rm SU}(3))=\mathbb{Z}$$
induced by the quotient procedure. In other words, if we take a homotopy
class $n \in \pi_3({\rm SU}(3))$ it corresponds to the $n$ modulo $4$ class in
$\pi_3(\mathcal{M}_3)$. 

We know also that the elements of $\pi_3({\rm SU}(3))$ 
are just  the ones of the embedded $\pi_3({\rm SU}(2))$.
The projection induced by the Hopf fibration gives 
a one-to-one correspondence between $$\pi_3({\rm SU}(2))=\mathbb{Z}\quad
{\rm and}\quad  
\pi_3(\mathcal{M}_2={\rm SU}(2)/{\rm SO}(2))=\mathbb{Z}\,.$$
This tells us that if we embed the solutions of the 
Faddeev--Skyrme model in $\mathcal{M}_3$, they are topologically stable
modulo 4. Thus, 
the solutions with the Hopf number $4 n $ are topologically trivial
in $\mathcal{M}_3$, while the others will tend to decay to 
the minimal $\mathbb{Z}_4$ representative.
This gives us an upper bound on the mass of each of the three
$\mathbb{Z}_4$ Skyrmions from the mass of the corresponding
Skyrmion in the Faddeev--Skyrme model (the three relevant ones are the
ones with the Hopf numbers $1$, $2$ and $-1$).

An interesting problem is to study the explicit
breaking of the  SU(3)  flavor symmetry in $\mathcal{M}_3$. 
To this end one can introduce a mass term $m_3\neq 0$ to the quark of the third
flavor in the microscopic theory. This   mass term breaks
the flavor group  SU(3) down  to  SU(2).
In the low-energy effective theory, with the chiral Lagrangian
(\ref{metric}), a potential term on $\mathcal{M}_3$ 
will be generated
(which will vanishes, of course, on the $\mathcal{M}_2$ submanifold).
If $m_3\to\infty$ all Skyrmion maps
are stable since $\pi_3 (\mathcal{M}_2)= \mathbb{Z}$. 

\begin{figure}[ht]
\begin{center}
\leavevmode
\epsfxsize 9  cm
\epsffile{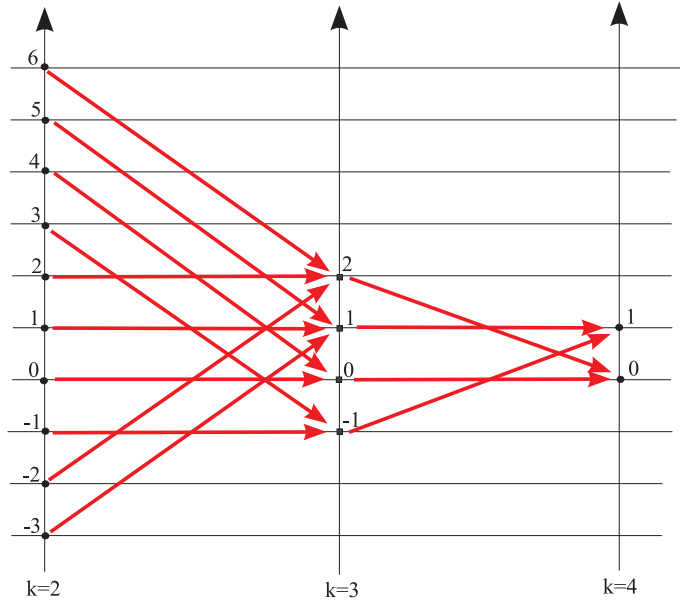}
\end{center}
\caption{\footnotesize The Skyrmions of the theory with $k=2$
(the Faddeev--Skyrme model)
are labeled by integer $n\in \mathbb{Z}$. If embedded in the theory with $k=3$,
they will tend to decay to the corresponding $\mathbb{Z}_4$
topological class. If we further embed the skyrmions in 
the theory with $k\geq 4$ only the Skyrmions with $\mathbb{Z}_2$
topological class will survive.
}
\label{decay}
\end{figure}

At finite $m_3$ the Skyrmions with the winding number 
larger than 2 and smaller than $-1$ will become metastable. They
will tunnel to the four stable configurations (see Fig.~\ref{decay}).
If $m_3$ is large enough, it should be possible   to calculate
the lifetimes of the metastable states
by using semiclassical methods.

If we further embed the model in $\mathcal{M}_k$
with $k \geq 4$, some of the $\mathbb{Z}_4$  
Skyrmions will become unstable and will decay into the
$\mathbb{Z}_2$ Skyrmions.

\section{Wess--Zumino--Novikov--Witten term}
\label{secfive}

If $\pi_4(G/H)$ is trivial there is no topological
obstruction for introduction of the WZNW term.\footnote{Ideas as to how one could introduce a nonstandard WZNW term in the cases of nontrivial $\pi_4(G/H)$ are discussed
in \cite{lmns}.}
This condition is satisfied for $k=3$ and for $k\geq 5$.
We can naturally generalize the expression from
the one referring to the SU$(N)$ case, discussed in Refs.~\cite{witten,witten2},
\beqn
S_{\rm WZNW} &\propto& \int_{B_5} d \Sigma^{\mu \nu \rho \sigma \lambda}\,
\Tr \,\Big\{ P_{h\bot}(U^\dagger \partial_\mu U)\cdot P_{h\bot}(U^\dagger \partial_\nu U)\nonumber\\[3mm]
&\cdot&  \! \! \! \! \!  P_{h\bot}(U^\dagger \partial_\rho U)\cdot P_{h\bot}(U^\dagger \partial_\sigma U)\cdot
 P_{h\bot}(U^\dagger \partial_\lambda U) \Big\} \,.
 \label{45}
\eeqn

In order to avoid an ambiguity in the quantization procedure 
due to different possible choices of $B_5$ for a given $S^4$ boundary
(see \cite{witten}),
we have to require the contribution of this term to be
a multiple of $2 \pi$  if integrated
on an arbitrary $S^5$ manifold.
The integral of this term over the $S^5$ manifold is a topological invariant
which depends on the topological class in $\pi_5(G/H)$.
The value of the integral vanishes for the $S^5$ cycles in finite cyclic
factors $\mathbb{Z}_k$ of $\pi_5(G/H)$ (there are indeed $\mathbb{Z}_2$
factors in $\pi_5(\mathcal{M}_k)$ for $k=3,5$, but they are irrelevant
for the WZNW term).  On the other hand, the integral can be non vanishing on
the $\mathbb{Z}$ factor, and its value is proportional to the ``winding number."

We have to normalize $S_{\rm WZNW}$ as follows:
\beqn  
S_{\rm WZNW} =n A && 
\int_{B_5} d\Sigma^{\mu \nu \rho \sigma \lambda}\,
\Tr \Big\{ P_{h\bot}(U^\dagger \partial_\mu U)\cdot P_{h\bot}(U^\dagger \partial_\nu U)
\nonumber\\[3mm]
 \cdot && 
 \! \! \! \! \! \!
 P_{h\bot}(U^\dagger \partial_\rho U) \cdot  P_{h\bot}(U^\dagger \partial_\sigma U) \cdot 
 P_{h\bot}(U^\dagger \partial_\lambda U) \Big\}\, ,
 \label{46}
 \eeqn
where the normalization factor $A$ is chosen in such a way that the integral on 
the map with the minimal winding
(in the $\mathbb{Z}$ factor of the $\pi_5$)
 between $S^5$ and $G/H$ is $2 \pi$
and $n$ is an arbitrary integer.

In the case of $\mathcal{M}_3$
we calculated  the WZNW term 
using the parameterization
introduced in Sect.~\ref{secthree}. 
It is proportional to the volume form of the manifold
(this is due to the fact that our target manifold is five-dimensional).
Namely, 
\beqn
S_{\rm WZNW} &=& n A\,  \frac{i \, 60}{64 \sqrt{3} }\,  \int_{B_5}\,  d \Sigma^{\mu \nu \rho \sigma \lambda}\, 
\left(  \partial_\mu  \theta \cdot \partial_\nu \eta \cdot
  \partial_\rho \alpha  \cdot\partial_\sigma \beta  \cdot\partial_\lambda \gamma
   \right )
  \nonumber\\[3mm]
 &\times&   \left(\cos \sqrt{3} \eta-\cos \theta\right) \sin \frac{\beta}{2}\,  \sin \theta\,.
   \label{47}
 \eeqn
The coefficient $A$ must be adjusted to make  the integral  $2 \pi$
on the map from $S_5$ to $\mathcal{M}_3$ corresponding to
the minimal winding. The element of 
$$\pi_5(\mathcal{M}_3)=\mathbb{Z} \times \mathbb{Z}_2$$ with 
the minimal winding in the $\mathbb{Z}$ factor
makes $l=2$ windings around the manifold. As a result we find the following
value for the normalization factor $A$:
\beq 
A=-\frac{2 i}{15 \pi^2}\,.
\label{48}
\eeq  

\section{Conclusions and Outlook}
\label{secsix}

If the fundamental quarks of QCD are replaced by massless adjoint
quarks the pattern of the chiral symmetry breaking is SU$(N_f)\to\,$SO$(N_f)$. 
This work addresses and solves the issue
of constructing sigma models on the coset spaces
${\rm  SU}(N_f)/{\rm SO}(N_f)$. The only  case which had been 
explicitly solved previously is $N_f =2$. This is the celebrated 
O(3) or CP(1) model.
We focused mainly on $N_f =3$,
presenting a full solution in this particular case,  with
a few general remarks on $N_f >3$ scattered in the bulk of the paper.
These remarks outline a general strategy for 
constructing the ${\rm  SU}(N_f)/{\rm SO}(N_f)$
sigma models for arbitrary $N_f$.

We  found an explicit parameterization for the sigma model
with the target space $\mathcal{M}_3={\rm SU}(3)/{\rm SO}(3)$ 
in terms of five angles.
The low-energy effective chiral Lagrangian is presented in Eqs.
(\ref{metric}), (\ref{47}) and (\ref{b2}).
As a check we computed the scalar curvature for the metric we got, 
and we found a constant, as is required for any homogeneous space.

We obtained WZNW term  too. Due to the fact 
$\mathcal{M}_3$ is a five-dimensional manifold,
the WZNW term  is proportional to the volume form.

We discussed the topological side of
the ${\rm  SU}(3)/{\rm SO}(3)$ sigma models. 
The nontrivial homotopy classes of $\pi_2(\mathcal{M}_3)=\mathbb{Z}_2$
and $\pi_3(\mathcal{M}_3)=\mathbb{Z}_4$, relevant for the vortex lines and  Skyrmions,
can be found by embedding in $\mathcal{M}_3$ 
some nontrivial homotopy classes of the Faddeev--Skyrme model. 

We can say that the algebraic aspect of the low-energy chiral dynamics
corresponding to the $\chi$SB pattern (\ref{doptue}) is in essence clear at the moment.
This problem has another aspect,  dynamical, related 
to interpreting the algebraic results obtained above in the language
of the underlying microscopic theory  --- Yang--Mills with the adjoint quarks.
Since $\pi_3 ({\mathcal M}_k)$ is nontrivial, the ${\rm  SU}(N_f)/{\rm SO}(N_f)$
chiral Lagrangians predict some ultraheavy stable solitons,
analogs of the QCD Skyrmions, whose mass scales as $N_c^2$
at large $N_c$. The question is can we understand these solitons
(and their stability) in the language
of the  microscopic (ultraviolet) theory?

This question obviously should be addressed and answered in the
framework of an independent  project whose thrust is
on dynamical roots of the soliton stability in the Yang--Mills theory with the adjoint quarks. The work in this direction has just started, with
first results reported in a follow-up publication \cite{BoShi}.

 In conventional QCD the Skyrme topological charge is matched with the
baryon number; in this way the Skyrmions can be identified with  baryons,
and their stability is protected by the global symmetry
--- the baryon charge conservation

 In adjoint QCD there is no such obvious reason for stability; 
 the analog of the baryon charge,
 the fermion number, is broken first to $\mathbb{Z}_{2 N_c N_f}$ 
by the chiral anomaly anomaly; this discrete symmetry is then spontaneously broken to $\mathbb{Z}_2$
by the fermion condensates.  This $\mathbb{Z}_2$ symmetry is not sufficient
by itself
to protect the soliton from decaying. This is due to the
fact that
in addition to the Goldstone bosons, which of course have vanishing fermion number,
 we expect 
light composite fermions of the form
\beq 
\psi_{\beta f} \propto \Tr(\lambda^\alpha_f \sigma^{\mu \nu}_{\alpha \beta} F_{\mu \nu}), 
\eeq
with an odd fermion number (in this expression $\lambda^\alpha_f$ is the adjoint Weyl fermion and
$\sigma^{\mu \nu}_{\alpha \beta} F_{\mu \nu}$ is the gluon field strength field in the spinorial notation). 

Reference \cite{BoShi} solves the problem of the soliton stability 
in the case  $N_f=2$. 
The solitons turn out to be in correspondence with exotic hadrons with 
mass $O(N_c^2)$ and $P=(-1)^Q(-1)^F=-1$, where $Q$ is the conserved charge corresponding to the 
unbroken $U(1)$ flavor subgroup. All other lighter degrees of freedom have $P=0$;
the Goldsone bosons have zero fermion number and even $Q$ charge; the light fermions $\psi$ have an
odd $Q$ charge and odd fermion number. This is just a $\mathbb{Z}_2$ stability
(a configuration with the Hopf number two can indeed decay to an array of $\pi$'s and $\psi$'s).
To detect this phenomenon in the low-energy  chiral theory we need
to introduce the fermions $\psi$ in the effective low-energy sigma-model.

This problem  for 
$N_f>2$ is currently under investigation.

\section*{Acknowledgments}

We are grateful to Stefano Bolognesi, Jarah Evslin and Andrei 
Losev for fruitful discussions.
This work   was
supported in part by DOE grant DE-FG02-94ER408.

\vspace{5mm}

\section*{Appendices}
\renewcommand{\theequation}{A.\arabic{equation}}
\setcounter{equation}{0}

\subsection*{A. Exact sequences for some homotopy groups}

\renewcommand{\thesubsubsection}{A.\arabic{subsubsection}}
\setcounter{subsubsection}{0}
\subsubsection{$\pi_2$}

The $k = 2$ case is special.
\[ \ldots  \rightarrow \pi_2\left({\rm SU}(k)\right)
\rightarrow 
 \pi_2\left({\rm SU}(2)/{\rm SO}(2)\right)\rightarrow \pi_1\left({\rm SO}(2)\right)
\rightarrow \pi_1\left({\rm SU}(2)\right)\rightarrow \ldots \]
\[ \ldots \rightarrow 0
\rightarrow 
 \mathbb{X}  \rightarrow \mathbb{Z}  \rightarrow 0
 \rightarrow \ldots, \]
which gives us $\mathbb{X}=\mathbb{Z}$.

For $k > 2$ we have the following exact sequence:
\[ \ldots  \rightarrow \pi_2\left({\rm SU}(k)\right)
\rightarrow 
 \pi_2\left({\rm SU}(k)/{\rm SO}(k)\right)\rightarrow \pi_1\left({\rm SO}(k)\right)
\rightarrow \pi_1\left({\rm SU}(k)\right)\rightarrow \ldots \]
\[ \ldots \rightarrow 0
\rightarrow 
 \mathbb{X}  \rightarrow \mathbb{Z}_2  \rightarrow 0
 \rightarrow \ldots, \]
which gives us $\mathbb{X}=\mathbb{Z}_2$.

\subsubsection{$\pi_3$}

For $k=2$ we know that the result is given by the Hopf fibration,
$\pi_3(S^2)=\mathbb{Z}$.

For $k = 3$ and $k \geq 5$ we have the following exact sequence:
\[ \ldots  \rightarrow \pi_3\left({\rm SO}(k)\right) \rightarrow \pi_3\left({\rm SU}(k)\right)
\rightarrow 
 \pi_3\left({\rm SU}(k)/{\rm SO}(k)\right)\rightarrow \pi_2\left({\rm SO}(k)\right)
\rightarrow \ldots \]
\[ \ldots \rightarrow \mathbb{Z} \rightarrow \mathbb{Z}
\rightarrow 
 \mathbb{X}   \rightarrow 0
 \rightarrow \ldots, \]
which gives us $\mathbb{X}=\mathbb{Z}_s$ where $s$
is the rank of the map between $\pi_3({\rm SO}(k))$ and
 $\pi_3({\rm SU}(k))$ induced by the embedding
 ${\rm SO}(k)\rightarrow {\rm SU}(k)$. 

The number $s$ can be calculated using the ``winding number"
integral discussed in Refs. \cite{bott,homotopy},
\beq 
s=-\frac{1}{24 \pi^2}  \int_{S^3} \Tr (U^\dagger dU)^3 \,, 
\eeq
where this integral is calculated on the smaller non-zero
element of $\pi_3({\rm SO}(k))$. 
For ${\rm SO}(3)$ a representative of the minimal element of $\pi_3$ is
\beq 
(\theta,\phi,\rho) \rightarrow \exp (i q_j \hat{n}_j \rho) \,,
\eeq
where $S^3$ is parameterized by
\[
 \hat{n}_j=(\sin \theta \cos \phi, \sin \theta \sin \phi,\cos \theta) \, ,\qquad 
  0<\rho<2 \pi \,,
 \]
and
\[ q_1= \pmatrix{  0    &  0   &    0
  \cr  0   & 0      & i
\cr  0    & -i     & 0 }, \quad 
 q_2= \pmatrix{  0    &  0   &    i
  \cr  0   & 0      & 0
\cr  -i    & 0     & 0 },\quad
 q_3= \pmatrix{  0    &  i   &    0
  \cr  -i   & 0      & 0
\cr  0    & 0     & 0 }. \]
For $k \geq 5$, we have to use
\[ q_1=1/2 \pmatrix{  0    &  0   &    i  & 0 & \ldots
  \cr  0   & 0      & 0  & -i & \ldots
\cr  -i    & 0     & 0   & 0 & \ldots
\cr  0    & i     & 0   & 0 &\ldots
\cr \vdots& \vdots&\vdots&\vdots&\ddots} , \quad
q_2=1/2 \pmatrix{  0    &  0   &   0 & i &\ldots
  \cr  0   & 0      & i  & 0 & \ldots
\cr  0    & -i     & 0   & 0 & \ldots
\cr  -i    & 0     & 0   & 0 &\ldots
\cr \vdots& \vdots&\vdots&\vdots&\ddots} ,\]
\[ q_3=1/2 \pmatrix{  0    &  i   &   0 & 0 &\ldots
  \cr  -i   & 0      & 0  & 0 &\ldots
\cr  0    & 0     & 0   & i &\ldots
\cr  0    & 0     & -i   & 0  & \ldots 
\cr \vdots& \vdots&\vdots&\vdots&\ddots},
 \] 
where the dots denote zeros. 
This gives $s=4$ for $k=3$ and $s=2$ for $k \geq 5$.

The $k = 4$ case is   particular,
\[ \ldots  \rightarrow \pi_3\left({\rm SO}(4)\right) \rightarrow \pi_3\left({\rm SU}(4)\right)
\rightarrow 
 \pi_3\left({\rm SU}(4)/{\rm SO}(4)\right)\rightarrow \pi_2\left({\rm SO}(4)\right)
\rightarrow \ldots \]
\[ \ldots \rightarrow \mathbb{Z} \otimes \mathbb{Z} \rightarrow \mathbb{Z}
\rightarrow 
 \mathbb{X}   \rightarrow 0
 \rightarrow \ldots. \]
Again the elements of  $\pi_3\left({\rm SU}(4)/{\rm SO}(4)\right)$
are in correspondence with the elements of $ \pi_3\left({\rm SU}(4)\right)$
which are not homotopic to any elements of  $ \pi_3\left({\rm SO}(4)\right)$. 
The same winding number argument used in the previous case 
for $k\geq5$ gives us $\mathbb{X}=\mathbb{Z}_2$. 

\subsubsection{$\pi_4$}

The $k = 2$ case is singled out,
\[ \ldots \rightarrow \pi_4\left({\rm SO}(2)\right) \rightarrow \pi_4\left({\rm SU}(2)\right)
\rightarrow 
 \pi_4\left({\rm SU}(2)/{\rm SO}(2)\right)\rightarrow \pi_3\left({\rm SO}(2)\right)
\rightarrow \ldots \]
\[ \ldots \rightarrow 0 \rightarrow \mathbb{Z}_2
\rightarrow 
 \mathbb{X}  \rightarrow 0 
 \rightarrow \ldots, \]
 which gives us $\mathbb{X}=\mathbb{Z}_2$.

For $k = 3$ and $k \geq 5$ we have the following exact sequence:
\[ \ldots  \rightarrow \pi_4\left({\rm SU}(k)\right)
\rightarrow 
 \pi_4\left({\rm SU}(k)/{\rm SO}(k)\right)\rightarrow \pi_3\left({\rm SO}(k)\right)
\rightarrow \pi_3\left({\rm SU}(k)\right)\rightarrow \ldots \]
\[ \ldots \rightarrow 0
\rightarrow 
 \mathbb{X}  \rightarrow \mathbb{Z}  \rightarrow \mathbb{Z}
 \rightarrow \ldots, \]
which gives us $\mathbb{X}=0$. 

The $k = 4$ case is also special,
\[ \ldots  \rightarrow \pi_4\left({\rm SU}(4)\right)
\rightarrow 
 \pi_4\left({\rm SU}(4)/{\rm SO}(4)\right)\rightarrow \pi_3\left({\rm SO}(4)\right)
\rightarrow \pi_3\left({\rm SU}(4)\right)\rightarrow \ldots \]
\[ \ldots \rightarrow 0
\rightarrow 
 \mathbb{X}  \rightarrow \mathbb{Z} \otimes \mathbb{Z}  \rightarrow \mathbb{Z}
 \rightarrow \ldots, \]
 which gives us $\mathbb{X}=\mathbb{Z}$.

\subsubsection{$\pi_5$}

The $k = 2$ case is special, as usual,
\[ \ldots \rightarrow \pi_5\left({\rm SO}(2)\right) \rightarrow \pi_5\left({\rm SU}(2)\right)
\rightarrow 
 \pi_5\left({\rm SU}(2)/{\rm SO}(2)\right)\rightarrow \pi_4\left({\rm SO}(2)\right)
\rightarrow \ldots \]
\[ \ldots \rightarrow 0 \rightarrow \mathbb{Z}_2
\rightarrow 
 \mathbb{X}  \rightarrow 0 
 \rightarrow \ldots, \]
 which gives us $\mathbb{X}=\mathbb{Z}_2$. 

For $k = 3,5$ we have the following exact sequence:
\[ \ldots  \rightarrow \pi_5\left({\rm SU}(k)\right)
\rightarrow 
 \pi_5\left({\rm SU}(k)/{\rm SO}(k)\right)\rightarrow \pi_4\left({\rm SO}(k)\right)
\rightarrow \pi_4\left({\rm SU}(k)\right)\rightarrow \ldots \]
\[ \ldots \rightarrow \mathbb{Z}
\rightarrow 
 \mathbb{X}  \rightarrow \mathbb{Z}_2  \rightarrow 0
 \rightarrow \ldots, \]
implying two alternatives, 
$$\mathbb{X}=\mathbb{Z}\quad
{\rm or} \quad \mathbb{X}=\mathbb{Z}\otimes \mathbb{Z}_2\,.$$ In Ref. \cite{rigas}
it is shown that the last option is the correct one. 

The $k = 4$ case is distinct,
\[ \ldots  \rightarrow \pi_5\left({\rm SU}(4)\right)
\rightarrow 
 \pi_5\left({\rm SU}(4)/{\rm SO}(4)\right)\rightarrow \pi_4\left({\rm SO}(4)\right)
\rightarrow \pi_4\left({\rm SU}(4)\right)\rightarrow \ldots \]
\[ \ldots \rightarrow \mathbb{Z}
\rightarrow 
 \mathbb{X}  \rightarrow \mathbb{Z}_2 \otimes \mathbb{Z}_2 \rightarrow 0
 \rightarrow \ldots, \]
  which gives us the alternatives $$\mathbb{X}=\mathbb{Z}\otimes \mathbb{Z}_2\quad
  {\rm or}  \quad \mathbb{X}=\mathbb{Z}\otimes \mathbb{Z}_2 
  \otimes \mathbb{Z}_2\,.
  $$
  It was shown n Ref.~\cite{rigas}
that the last choice is the correct one.

For $k = 6$ we get
\[ \ldots \rightarrow \pi_5\left({\rm SO}(k)\right) \rightarrow \pi_5\left({\rm SU}(k)\right)
\rightarrow 
 \pi_5\left({\rm SU}(k)/{\rm SO}(k)\right)\rightarrow \pi_4\left({\rm SO}(k)\right)
\rightarrow \ldots \]
\[ \ldots \rightarrow \mathbb{Z} \rightarrow \mathbb{Z}
\rightarrow 
 \mathbb{X}  \rightarrow 0 
 \rightarrow \ldots, \]
 which is not enough to find $\mathbb{X}$.
  In Ref. \cite{rigas}
it was shown that $\mathbb{X}=\mathbb{Z}$. 

For $k > 6$:
\[ \ldots \rightarrow \pi_5\left({\rm SO}(k)\right) \rightarrow \pi_5\left({\rm SU}(k)\right)
\rightarrow 
 \pi_5\left({\rm SU}(k)/{\rm SO}(k)\right)\rightarrow \pi_4\left({\rm SO}(k)\right)
\rightarrow \ldots \]
\[ \ldots \rightarrow 0 \rightarrow \mathbb{Z}
\rightarrow 
 \mathbb{X}  \rightarrow 0 
 \rightarrow \ldots, \]
 which gives $\mathbb{X}=\mathbb{Z}$. 

\subsection*{B. Four-derivatives term}
\renewcommand{\theequation}{B.\arabic{equation}}
\setcounter{equation}{0}

The four-derivatives term can be computed from Eq.~(\ref{la4}).
Let us introduce the following compact notation:
\beq 
S^{\mu \nu}_{(\theta,\eta)}=
\partial_\mu \theta \,\,   \partial_\nu \eta - \partial_\mu \eta \, \,  \partial_\nu \theta \,, 
\label{b1}
\eeq
and the same for all other possible
coordinate pairings among  $\theta, \eta, \alpha, \beta, \gamma$.
Then we obtain, after a rather straightforward but quite cumbersome calculation, the following explicit expression:
\beqn
&& \mathcal{L}_4=8 \sin^2 \theta   (S^{\mu \nu}_{(\theta,\alpha)})^2 
\nonumber\\[2mm]
&& +(1-\cos \sqrt{3} \eta \cos \theta-\cos \alpha \sin \sqrt{3} \eta \sin \theta) (S^{\mu \nu}_{(\theta,\beta)})^2\nonumber\\[2mm]
&&+\sin^2 \theta
(1-\cos \sqrt{3} \eta \cos \theta-\cos \alpha \sin \sqrt{3} \eta \sin \theta) (S^{\mu \nu}_{(\alpha,\beta)})^2\nonumber\\[2mm]
&&+3 (1-\cos \sqrt{3} \eta \cos \theta-\cos \alpha \sin \sqrt{3} \eta \sin \theta)(S^{\mu \nu}_{(\eta,\beta)})^2 \nonumber\\[2mm]
&&+3 \sin^2 \frac{\beta}{2} (1-\cos \sqrt{3} \eta \cos \theta+\cos \alpha \sin \sqrt{3} \eta \sin \theta)
 (S^{\mu \nu}_{(\gamma,\eta)})^2 \nonumber\\[2mm]
&& +\frac{1}{2}\left[ 8 \sin^2 \theta +(1-\cos \sqrt{3} \eta \cos \theta+\cos \alpha \sin \sqrt{3} \eta \sin \theta)  \right.\nonumber\\[2mm]
&& \left. + \cos \beta (8 \sin^2 \theta-(1-\cos \sqrt{3} \eta \cos \theta
+\cos \alpha \sin \sqrt{3} \eta \sin \theta)  )
  \right] (S^{\mu \nu}_{(\theta,\gamma)})^2 \nonumber\\[2mm]
&& + \sin^2 \theta \sin^2 \frac{\beta}{2} 
 (1-\cos \sqrt{3} \eta \cos \theta+\cos \alpha \sin \sqrt{3} \eta \sin \theta)
 (S^{\mu \nu}_{(\alpha,\gamma)})^2\nonumber\\[2mm]
&&+\frac{1}{8} \left[ 4 -\cos \alpha \sin \sqrt{3} \eta \sin^3 \theta 
 -\cos \alpha \cos \beta \sin \sqrt{3} \eta \sin^3 \theta+
 3 \cos \beta \sin^2 \theta + \sin^2 \theta  \right.\nonumber\\[2mm]
&&-3 \cos \alpha \sin \sqrt{3} \eta \sin \theta 
 -3 \cos \alpha \cos \beta \sin \sqrt{3} \eta \sin \theta -
 (\cos \beta-1) \cos^2 \sqrt{3} \eta
 \nonumber\\[2mm]
&& + \cos \beta \sin^2 \sqrt{3} \eta - \sin^2 \sqrt{3} \eta -
 \cos \sqrt{3} \eta \cos \theta (6-2 \cos \beta-4 \cos^2 \frac{\beta}{2} \cos 2 \theta) 
 \nonumber\\[2mm]
&& \left. - \cos^2 \theta \left(1-3 \cos \alpha \sin \sqrt{3} \eta \sin \theta + \cos \beta
 (3-3 \cos \alpha \sin \sqrt{3} \eta \sin \theta)  \right)  \right] (S^{\mu \nu}_{(\beta,\gamma)})^2 
 \nonumber\\[2mm]
&& - 2 \sin \alpha \sin \frac{\beta}{2} \sin \sqrt{3} \eta \sin^3 \theta 
 (S^{\mu \nu}_{(\beta,\alpha )})   (S^{\mu \nu}_{(\gamma,\alpha )})  
 \nonumber\\[2mm]
&& + 2 \cos \frac{\beta}{2}  \sin^2 \theta 
 \left(1-\cos \sqrt{3} \eta \cos \theta-\cos \alpha \sin \sqrt{3} \eta \sin \theta\right)
 (S^{\mu \nu}_{(\beta,\alpha )})   (S^{\mu \nu}_{(\beta,\gamma )})  
 \nonumber\\[2mm]
&& + \sin \alpha \sin \beta \sin \sqrt{3} \eta \sin^3 \theta
 (S^{\mu \nu}_{(\gamma,\alpha )})   (S^{\mu \nu}_{(\gamma,\beta )}) 
 \nonumber\\[2mm]
&&-2 \sqrt{3} \cos \frac{\beta}{2} \sin \alpha \sin \theta  (\cos \sqrt{3} \eta-\cos \theta) 
 (S^{\mu \nu}_{(\gamma,\beta )})   (S^{\mu \nu}_{(\eta,\beta )}) 
 \nonumber\\[2mm]
&&- \sqrt{3} \cos \alpha \sin \beta \sin \theta  (\cos \sqrt{3} \eta-\cos \theta) 
 (S^{\mu \nu}_{(\gamma,\beta )})   (S^{\mu \nu}_{(\gamma,\eta )}) 
 \nonumber\\[2mm]
&&- 6 \sin \alpha \sin \frac{\beta}{2} \sin \sqrt{3} \eta \sin \theta
 (S^{\mu \nu}_{(\eta,\beta )})   (S^{\mu \nu}_{(\eta,\gamma )}) 
 \nonumber\\[2mm]
&&+ 2 \sqrt{3}  (\cos \sqrt{3} \eta-\cos \theta) \sin \alpha \sin^2 \frac{\beta}{2}
\sin \theta  (S^{\mu \nu}_{(\gamma,\alpha )}) (S^{\mu \nu}_{(\gamma,\eta )}) 
 \nonumber\\[2mm]
&&- 2 \sqrt{3}  (\cos \sqrt{3} \eta-\cos \theta) \sin \alpha 
\sin \theta  (S^{\mu \nu}_{(\beta,\alpha )}) (S^{\mu \nu}_{(\beta,\eta )}) 
  \nonumber\\[2mm]
&& - 3 (\cos \sqrt{3} \eta-\cos \theta) \sin \beta \sin \theta
(S^{\mu \nu}_{(\gamma,\beta )}) (S^{\mu \nu}_{(\gamma,\theta )})    
\nonumber\\[2mm]
&&- 2 \sin \alpha \sin \frac{\beta}{2} \sin \sqrt{3} \eta \sin \theta
(S^{\mu \nu}_{(\theta,\beta )}) (S^{\mu \nu}_{(\theta,\gamma )})  
\nonumber\\[2mm]
&& +16 \cos \frac{\beta}{2} \sin^2 \theta  
(S^{\mu \nu}_{(\theta,\alpha )}) (S^{\mu \nu}_{(\theta,\gamma )}) 
\nonumber\\[2mm]
&&- 2 \sqrt{3} \sin^2 \frac{\beta}{2}(
 \cos \alpha ( \cos \sqrt{3} \eta \cos \theta -1) -\sin \sqrt{3}\eta \sin \theta )
 (S^{\mu \nu}_{(\eta,\gamma )}) (S^{\mu \nu}_{(\theta,\gamma )})
 \nonumber\\[2mm]
&&+ 2 \sqrt{3} (
 \cos \alpha ( \cos \sqrt{3} \eta \cos \theta -1) +\sin \sqrt{3}\eta \sin \theta )
 (S^{\mu \nu}_{(\eta,\beta )}) (S^{\mu \nu}_{(\theta,\beta )}) 
 \nonumber\\[2mm]
&&+4 \sqrt{3} (\cos \sqrt{3} \eta \cos \theta -1) \sin \alpha \sin \frac{\beta}{2} 
 (S^{\mu \nu}_{(\eta,\gamma )}) (S^{\mu \nu}_{(\theta,\beta )}) 
 \nonumber\\[2mm]
&& +2 \sqrt{3} (\cos \sqrt{3} \eta \cos \theta -1) \sin \alpha \sin \frac{\beta}{2} 
 (S^{\mu \nu}_{(\gamma,\beta )}) (S^{\mu \nu}_{(\theta,\eta )})
 \nonumber\\[2mm]
&&
- 6 (\cos \theta - \cos \sqrt{3} \eta) \sin \frac{\beta}{2} \sin \theta
 (S^{\mu \nu}_{(\gamma,\beta )}) (S^{\mu \nu}_{(\theta,\alpha )})
 \nonumber\\[2mm]
&&- 4 \sqrt{3} \sin \theta (\cos \sqrt{3}  \eta -\cos \theta ) \cos \alpha \sin \frac{\beta}{2} 
 (S^{\mu \nu}_{(\gamma,\alpha )}) (S^{\mu \nu}_{(\eta,\beta )})
 \nonumber\\[2mm]
&&- 2 \sqrt{3} \sin \theta (\cos \sqrt{3}  \eta -\cos \theta ) \cos \alpha \sin 
\frac{\beta}{2} 
 (S^{\mu \nu}_{(\gamma,\beta )}) (S^{\mu \nu}_{(\alpha,\eta )})\,.
 \label{b2}
\eeqn

\end{document}